\documentstyle[pre,aps,multicol,epsfig,color]{revtex}
  
\begin{document}
\draft     

\title{Consequence of reputation in an open-ended Naming Game} 

\author{Edgardo Brigatti }  
  
\address{Instituto de F\'{\i}sica, Universidade Federal Fluminense, 
  Campus da Praia Vermelha, 24210-340, Niter\'oi, RJ, Brazil}
\address{Centro Brasileiro de Pesquisas F\'isicas,
Rua Dr. Xavier Sigaud 150,
22290-180, Rio de Janeiro, RJ, Brazil}
\address{e-mail address: edgardo@if.uff.br}

\date{\today}
\maketitle
\widetext
  
\begin{abstract}
 
We study a modified version of the Naming Game, a recently 
introduced model which describes
how shared vocabulary can 
emerge spontaneously in a population without any central control.  
In particular, we introduce a new mechanism that allows a 
continuous interchange with the external inventory of words. 
A novel  playing strategy, influenced by the hierarchical 
structure that individuals' reputation defines in the community, is implemented.
We analyze how these features influence the convergence times, 
the cognitive efforts of the agents and the scaling behavior in 
memory and time. 

\end{abstract}
  

\begin{multicols}{2}

\section{Introduction}

The origin and spread of languages and the 
evolution of their differentiation are problems addressed by 
various theories that cross different philosophical orientations, 
varying from 
nativism
and evolutionary approaches 
to behaviorism and conventionalism.
A good archetype of this last perspective can 
be retrieved in the last of Wittgenstein's reflections \cite{witt}.
There, language is seen as an activity
that arbitrarily attributes meanings to words
throughout the function they assume in the life
of humans. 
Meaning is defined by the use of language: `` 
the meaning of a word is its use in the language \cite{witt} ''.
In this light, language is a sort of training to react in a specific way 
in relation to a specific sign:
 a {\it language game}. 

A particular linguistic problem that can be considered as a 
grounding test for this hypothesis is the rise of a new linguistic quantity.
Linguists tried to characterize in a
quantitative way such 
changes \cite{best}, using simple mathematical models to 
describe the rise (or fall) of 
a linguistic element \cite{vulanovic}. 
In particular, among the possible different linguistic changes,
we will point our attention to learning processes characterized by 
fast dynamics, as, for example, the birth of neologisms. 
Looking at dictionaries, 
it is possible to see how every year thousands of 
new words appear or substitute others. Moreover, reading and comparing throughout
different periods some newspapers, we can observe how many words 
or syntactic changes spread out or substitute the old ones. 
Finally, we can observe the emergence and death of jargon, technical 
words or idiomatic expressions. 
These facts may be considered as a good paradigm 
for testing more general theories, which can also account for long-term processes, 
in the same way as, in biology, the study of very fast evolutionary shifts, 
comparable to the life-span of a human being (microevolution),
can give insights into the behavior of
evolutive processes characterized by geological time-scales.\\

A well known artificial experiment implemented to simulate these fast learning dynamics, 
with the aim of testing the general hypothesis concerning the origin of languages which 
we have sketched above, is the ``Talking Heads Experiment'' \cite {steels1}.
There, embodied software agents bootstrap a shared lexicon 
without any external intervention. Robots concretize a {\it language game} 
developing a vocabulary throughout a self-organized process, a Naming Game. 
In resemblance with the Wittgensteinian point of view, language can be seen 
as an autonomous adaptive system shaped and reshaped by the use and the 
behavior of the local linguistic activity \cite {steels2}.
   
Recently, these studies have also attracted the interest of the 
Statistical Physics community.
The dynamics of such Naming Games is characterized by a period of spread and 
diffusion of new competing words, followed by a sudden transition \cite{best,vulanovic} 
towards the use of a single word.  These facts are quite common
to other well known social dynamics \cite{sociophysics}, where a population aims to reach a 
common and shared state, the consensus \cite{socioconsensus}.
One of the novelties of these studies consists in the fact of abandoning any 
evolutionary approach \cite{evo}, dealing with the emergence of communication 
conventions on fast time-scales. Moreover, no central control, that can determine 
a global coordination, is considered, even in the form 
of some selection force. \\

A first study in this direction has recently appeared \cite{baronca}, 
directly inspired by the experiments conducted with the use of robots \cite {steels2}.
In that work the Naming Game is modeled as simply as possible, with the aim of 
implementing a lower bound in complexity and processing power.
Each player is characterized by an inventory of words associated with 
an object.
At each time step two players, randomly chosen, interact following 
some simple rules.
The speaker retrieves a word from his inventory, or, if his inventory 
is empty, invents a new word and transmits the selected word to the hearer. 
If the hearer's inventory contains such 
a word the communication is a success. The two agents update their inventories so 
as to keep only the word involved in the interaction.
Otherwise, the communication is a failure and the hearer adds an association 
between the new word and the object.
These simple rules put into action three mechanisms: an 
uploading mechanism that introduces new words from an external 
inventory of words, an overlapping mechanism
that allows the spreading of a particular word among the players 
and an agreement mechanism that deletes useless words.  
With these simple mechanisms the system undergoes a disorder/order transition 
towards an absorbing state characterized by a single word for 
all the players. This behavior scales-up to very large populations.\\

As stated before,  Baronchelli et al. implemented their Naming Game 
inspired by 
the behavior of the Talking Heads experiment. 
In this paper, we are interested in 
 modeling some features more related to a real community of speakers, such as for example a 
classroom of pupils becoming competent in a language, a community of foreigners learning a 
new language or the dynamics of jargon creation in 
a metropolitan tribe or in a group of researchers coping with new objects or concepts. 
Looking at these situations, the real world language is open-ended,
with no evident constraints on the possible number of different words.
We can evidence a sort of fluidity by which new words can enter 
or leave the lexicon inventory.

In contrast, in the original model \cite{baronca}, each agent can store an 
unlimited number of different names only potentially, and not as a matter 
of fact . 
This may be understood if we look with attention to that dynamics. 
Even if the agents can store an unlimited number of words, 
the rules of the Game allow the introduction of 
new words only if the agent's inventory is empty, 
and this happens just in the first Monte Carlo steps.
After this fast transient, when everybody has, at least, one word, 
the system manifests itself like a closed system and no other new words are included. 
The Game is characterized by a fixed number of different words that, throughout
the overlapping and agreement mechanism, reduces to one.

A second point that we want to investigate is the limited feedback 
between speakers
in the case of a failed communication and its relation with social structures. 
In a real situation of failure, the hearer is led to learn 
the association between the object and the word only if he 
recognizes the speaker as a sort of 
teacher. Even the fact that the speaker and the hearer are able to establish, 
by means of a subsequent action, if the talk was successful or not,
does not seem a sufficient factor to justify the learning of a new association.
In contrast, in the original definition of the Naming Game, 
the overlapping mechanism always forced 
the hearer to learn from the speaker. 
This dynamics is perhaps the most powerful for
reaching the consensus state 
but may be considered realistic only in the case where the speaker
has the fixed role of monitoring and the hearer of reproducing.
In other words, if the speaker acts as a teacher and the 
hearer like a student.
It is realistic to suppose that, in general, these
roles are defined by the social structure of the community, and not
randomly assigned during each communication.

The previous scenario, described by Baronchelli et al.  \cite{baronca}, 
assumed players operating under full anonymity.  
A general attention for the social structure of the community and
for the role of population heterogeneity has already 
appeared in later works, which defined heterogeneous topologies,
where different agents play different roles \cite{luca},
or non-completely random interactions  \cite{amichetti}.
However, in our work, we want to point our attention to 
more specific facts.
In situations relating to humans, players may 
accumulate information about their 
environment and specifically about potential future interaction 
partners. 
All players carry some sort of reputation reflecting 
their success in communicate and, 
through observation
of third-party interactions and gossip, a player's 
reputation may become known to others.
Finally, 
we can suppose that all players care for 
learning from agents known
to generally obtain successful interactions.\\


We were naturally led to explore if the model of Baronchelli et al. 
is robust to changes that contemplate these general assumptions
and what effects these elements have on the dynamics and statistical 
behavior of the system.
We are interested in introducing these new elements not only to  
describe a more realistic situation, but to test the 
robustness of the mechanisms outlined in the previous model
and to investigate if some simple new structure is able to improve
our system performance in reaching consensus.

For these reasons, in the remainder of the paper, we will present our 
new version of the model. We will describe an  
open system where each agent can actually store an unlimited number 
of different names. This  fact is possible thanks to a dynamics that 
allows 
the introduction of different words at every M.C. step.
Moreover, we introduce a  hierarchical structure between the 
agents playing our Naming Game, making it possible to  
distinguish between the players which 
act as teachers and the ones which act as 
learners. This will be obtained through the establishment of
the concept of status or reputation, a universal feature 
of human sociality, that can be generally related to
numerous large-scale human collective
behaviors \cite {alexander}.
The importance of introducing this concept becomes clear
if we look at language as a kind of 
collective, and not individualistic, problem solving process.


\section{The model}

The Game is played by $P$ agents. An inventory that can 
contain an arbitrary number of words represents each agent.
Moreover, an integer number ($R$) labels each player and represents 
its reputation across the community. 
We introduce reputation as a score \cite{nowak} which is 
variable in time. 
The population starts with a random distribution of the $R$
values, and during the time evolution the reputation of
each player changes according to its performance during 
the game, following the rules explained below.
During each interaction, the agent with 
greater reputation acts like a teacher, and the other 
one like a learner.\\ 

At each time step, the following microscopic rules control our model:

1) The speaker, with reputation $R_{S}$, retrieves a word 
from its inventory or, if its inventory is empty, invents a new word.

2) The speaker transmits the selected word to the hearer, characterized by
the reputation $R_{H}$.

3a) If the hearer's inventory contains such a word, 
the communication is a success. 
The two agents update their inventories so as to keep only the word involved in the 
interaction. The speaker's reputation increases by one.
 
3b) Otherwise the communication is a failure. 
If $R_{S}>R_{H}$, the hearer adds the new word to its inventory
and the speaker does nothing. 
If $R_{S}<R_{H}$, the speaker invents a new word and the 
hearer does nothing.
The speaker's reputation decreases by one.\\


The implementation of these rules 
defines an open-ended system where an unlimited number of words
can be really invented.
Players invent new words if their inventory is empty
(that happens only in the early stages of the simulation),
or if their communication is a failure. In fact, we can think that,
in real life, individuals that are not able to communicate are naturally 
led to look for new words. 
 
The process determines, across the population, 
a hierarchical structure that allows one to 
define distinct roles during each communication 
event.  This structure is dynamical and changes throughout 
the temporal evolution.
Every player is continually assessed: reputation is defined as a score, 
that can be high or low, depending 
on the previous rounds of the play.

\section{Numerical results}

We will describe the time evolution of our system looking at some  
usual global quantities \cite{baronca}: the total number of words ($N_{tot}$) 
present in the population, the number of different words ($N_{dif}$) 
and the success rate ($S$), that measures an average rate of success 
of communications.

\begin{figure}[p]
\begin{center}
\includegraphics[width=0.45\textwidth, angle=0]{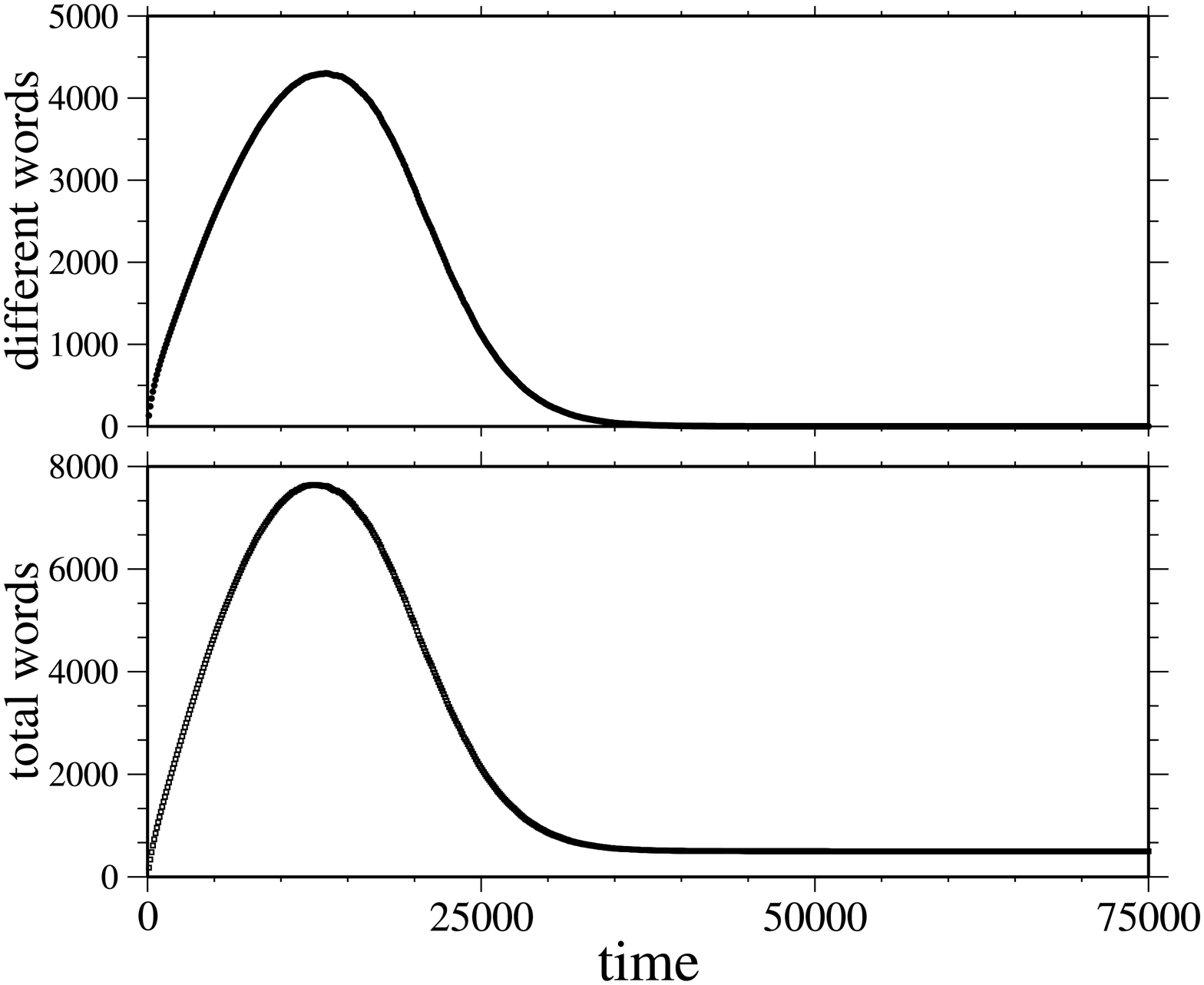}
\includegraphics[width=0.45\textwidth, angle=0]{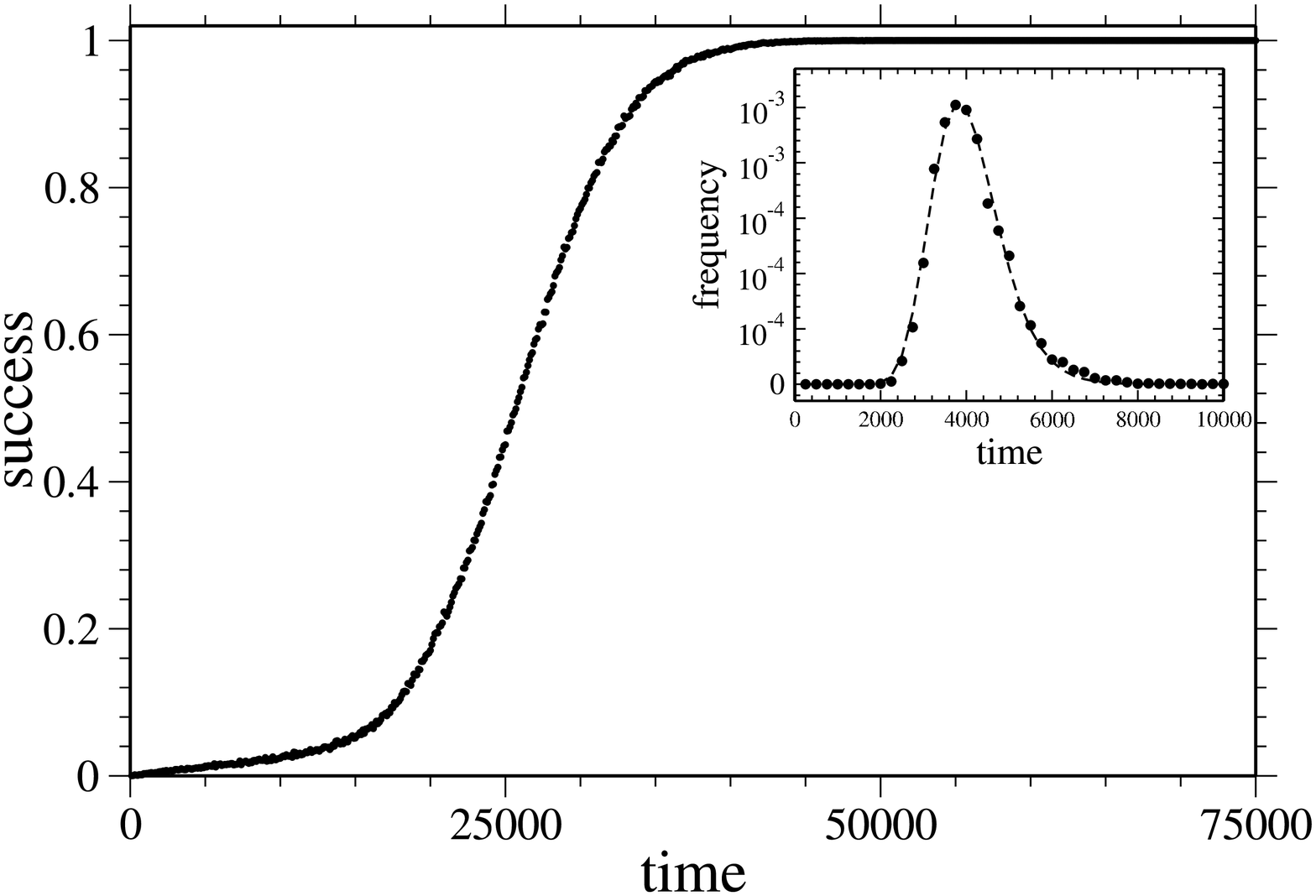}
\end{center}
\caption{\small Top: temporal evolution for the total number of words ($N_{tot}(t)$) 
and for the number of different words ($N_{dif}(t)$). 
Bottom: the success rate ($S(t)$). 
Data are averaged over 100 simulations with $P=500$. 
The agents' initial reputations
follow a Gaussian distribution with 
standard deviation $5$. 
In the inset we present the distribution of the
convergence time ($P=100$, $\sigma=5$).
Data are fitted by a lognormal distribution.
}
\label{fig_pheno}
\end{figure}

An initial transient exists where agents have an empty inventory. 
In this early stage, in each interaction, each speaker invents at least 
a new word and each hearer can possibly learn one.
Already in this phase, the behavior of our model differs from the one of the 
original Naming Game. In fact, in those simulations, 
this phase corresponds to the rise of $N_{dif}$
and finishes when the total number of different words 
reaches its maximum, equal to $N/2$, that
is maintained along a plateau.  
With our model, the curves $N_{tot}(t)$ and $N_{dif}(t)$ behave 
in the same manner and constantly increase until a maximum is 
reached at $t_{max}$. During this long learning phase, the total 
number of distinct words does not display any plateau 
(see Figure \ref{fig_pheno}). 

When the redundancy of words reaches a sufficiently high level, 
the number of successful plays increases. 
The curves $N_{tot}(t)$ and $N_{dif}(t)$ begin a decay 
towards the consensus state, corresponding to 
one common word for all the players, reached at time $T_{con}$.

In these dynamics it is possible to distinguish between two phases.
A first one, where the system reorganizes itself, building correlations 
as a consequence of a collective behavior. It starts when the time 
evolution reaches $T_{max}$ and $S(t)$ maintains 
a linear increase in time. A second one, when the disorder/order transition   
takes place and a very fast convergence process  toward the absorbing state 
occurs. Also in our model, the system passes through a quick reorganization 
before entering the fast transition dynamics.
In Figure \ref{fig_pheno} we report the temporal evolution for $N_{tot}(t)$, 
$N_{dif}(t)$ and $S(t)$. In these simulations 
the initial values of the agent's reputation follow
a Gaussian distribution centered in $0$ with standard deviation $\sigma=5$.\\

\begin{figure}[p]
\begin{center}
\includegraphics[width=0.45\textwidth, angle=0]{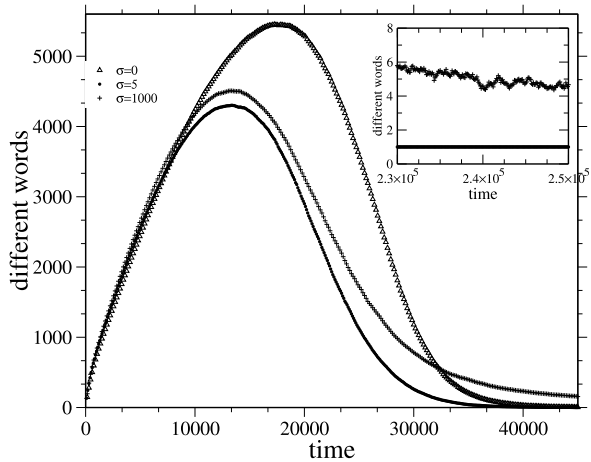}
\end{center}
\caption{\small Temporal evolution for the number of different words 
for population with a reputation obtained from Gaussian 
distributions with different standard deviations ($\sigma$).
The inset shows the very slow 
convergence towards the absorbing state,
characterized by the presence of  long standing 
quasi-stationary states.
Data averaged over 100 simulations with $P=500$.} 
\label{fig_hierar1}
\end{figure}

In general, the system evolution is strongly dependent on the 
initial condition of the reputations' distribution. 
The evolution is obviously not dependent on the mean value of the $R$ 
distribution, but depends on the value of its spread $\sigma$. 
For example, if we start with every player characterized by the same 
value of $R$, 
for the same parameters, the maximum of the total number of words reaches 
the largest value. In other words, in this situation the system needs 
the greatest memory size. 
Increasing the spread value the necessary memory size decreases, 
until reaching a minimal value. 
For instance, if we chose the $R$ values from a Gaussian distribution, 
and  we increase the standard deviation,
the maximal memory necessary to begin producing successful plays decreases, 
reaching a minimal value for a standard deviation equal to $5$ 
(an optimal value that is preserved for different population size). 
In this interval of $\sigma$ values,
the convergence time towards consensus seems to be not very different. 
In contrast, if we further increase the spread, 
$T_{con}$ considerably increases. 
Long standing quasi-stationary states appear, 
characterized by the presence of  
a fixed small number of words designating the same object.  
Figure \ref{fig_hierar1} clearly shows this behavior.

We can understand this fact noting that, if an agent has a relatively
high reputation, in general,  it will not learn words from 
other players. 
On the other side,
it will have a greater chance to propagate its own words.
Agents with the highest $R$ value will act as nucleation 
points, causing the spreading of the words that generally survive in the
final state. For this reason, in population with a broad distribution in the 
reputation, an higher number of prestigious agents form different clusters 
of agents with the same words and a coarsening dynamics generates a very 
slow evolution towards consensus. \\

\begin{figure}[p]
\begin{center}
\includegraphics[width=0.45\textwidth, angle=0]{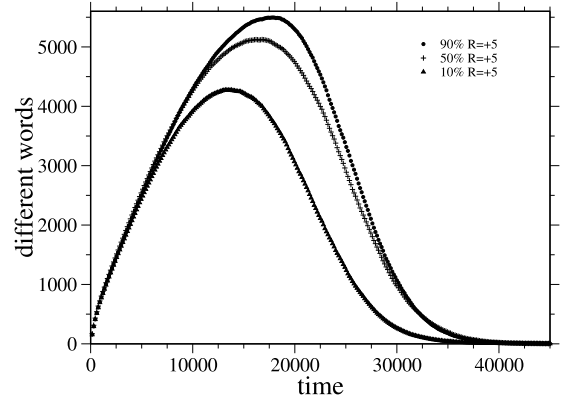}
\end{center}
\caption{\small Temporal evolution for the number of different words 
for populations with an initial asymmetric 
distributions in the $R$. 
The population with $10\%$  of agents having $R=+5$ (teachers), 
and the others having $R=-5$,  shows the minor memory 
requirement. 
The population with $90\%$ of teachers needs the 
largest amount of memory.
Data averaged over 100 simulations with $P=500$.} 
\label{fig_hierar2}
\end{figure}

Furthermore, we explore how initial asymmetric 
distributions of $R$ could affect the system evolution 
(see Figure \ref{fig_hierar2}).
We run simulations with
two classes of agents: one with $R=5$ and the other with  $R=-5$.
When the sub-population of $R=5$ players is majority, 
the system performance, in terms of memory cost, 
gets worst . 
In contrast, if the number of players with
a high reputation is smaller, the memory cost necessary for reaching 
consensus is sensibly reduced.
These results have a simple interpretation. It is easier to reach consensus 
in an authoritarian community, where the few individuals with a high 
reputation can easily and efficiently spread their words among population.

\section{ Scaling laws}

From the viewpoint of applications and for better understanding the 
model behavior we investigate how the 
macroscopic observables scale with the size $P$ of the population.

\begin{figure}[p]
\begin{center}
\includegraphics[width=0.45\textwidth, angle=0]{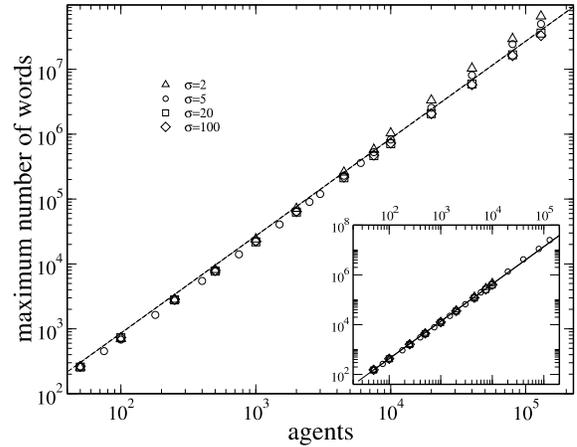}
\end{center}
\caption{\small Maximum number of total words and, in the inset,
maximum number of different words for different population sizes.
The lines have slope $3/2$.
In all simulations the agents' initial reputations
follow a Gaussian distribution.
These results are robust with respect to the choice of
different  $\sigma$ values.} 
\label{fig_maxw}
\end{figure}

At first, we look at the scaling behavior of the system memory size.
The maximum number of total words ($Max[N_{tot}]$) and
the maximum number of different words scale according to the
same power law: $P^{3/2}$ (see Figure \ref{fig_maxw}). 
The behavior of $Max[N_{tot}]$ is the same founded by 
Baronchelli et al. \cite{baronca}.
In contrast, in their work, the maximum number 
of different words scales as $P$,
because it is governed by a different dynamics that does not allow 
the introduction of an unlimited number of new different words.

\begin{figure}[p]
\begin{center}
\includegraphics[width=0.45\textwidth, angle=0]{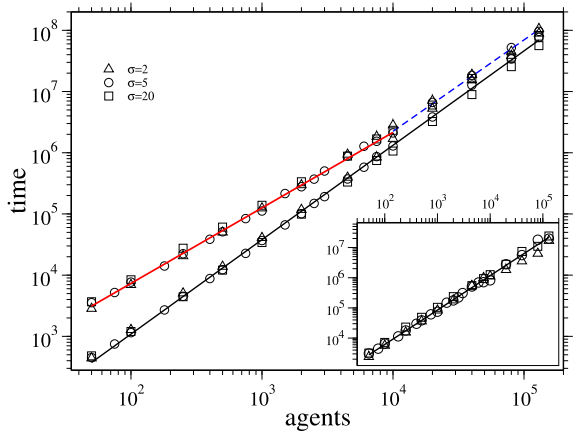}
\end{center}
\caption{\small  Convergence time ($T_{con}$, data on the top) and time position of the peak 
that corresponds to the maximum number of total words ($T_{max}$, data on the bottom) 
for different population sizes $P$.
$T_{con}$ displays a crossover: for $P<10000$  $T_{con}\propto P^{1.23\pm0.01} $,
for $P>10000$  $T_{con}\propto P^{1.49\pm0.03} $.
$T_{max}$ is well described by a power law with 
exponent $ 1.54\pm0.01$.
The inset shows the power law dependence of the 
alignment  time $T_{con}-T_{max}$  (the fitted 
exponent is $1.14\pm0.02$).
We present simulations where the agents' initial 
reputations follow a Gaussian distribution with different 
$\sigma$ values.}
\label{fig_tconv}
\end{figure}

In our model, the scaling of the time position of the peak that corresponds 
to the maximum number of total words is: $T_{max}\propto P^{3/2}$ 
(see Figure \ref{fig_tconv}).
To sum up, the dynamics of accumulation and
spread of words of our model is very similar 
to the one given by the model described in \cite{baronca},
which required a large agents' memory ($P^{3/2}$) 
and a long period of words exchanging between players ($P^{3/2}$).
A way of improving this behavior, experimented 
for the model in \cite{baronca}, is its
implementation on a low dimensional regular lattice,
where a minor memory requirement is found, 
at the cost of a very slow convergence towards 
consensus \cite{baronca1}.\\

The time necessary to reach convergence to the global consensus 
($T_{con}$) is the other fundamental quantity that characterizes 
our system.
$T_{con}$ displays an interesting behavior: for small communities ($P<10000$) 
the hierarchical structure defined by the reputation parameter has a strong 
influence over convergence and:
\begin{eqnarray} 
\label{eq1}
T_{con}\propto P^{1.2}\nonumber
\end{eqnarray} 

This scaling behavior is slower than the one found 
by Baronchelli et al. \cite{baronca},
where $T_{con}\propto P^{3/2}$. It is also slower of 
the same model when embedded 
in a small-world topology, where the convergence process has a 
$P^{1.4}$-dependence  \cite{asta} (we remind the reader that  slower scaling behavior in $P$ means  faster convergence time).
Unfortunately, the positive effect of reputation breaks up as the dimension of the 
community grows.
For a population larger than $10000$ individuals there is a 
sudden change
(see Figure \ref{fig_tconv}) and the new dependence becomes  $P^{3/2}$. 
This large scale behavior of $T_{con}$ is an expected outcome: 
in all the implementations of these models, 
the $P$ dependence of $T_{con}$ must be higher or 
equal to  the $T_{max}$ dependence. 
From this perspective, the temporal scaling behavior of the 
dynamics of accumulation 
and spread of words ($T_{max}$), which usually scales as  
the system memory size, directly influences the $P$ dependence of $T_{con}$.

We can better understand our results with the help of some simples considerations.
The convergence time encompasses the time necessary for the conclusion of two different
dynamical processes which can have distinct scaling behaviors. 
The first process, which occurs from the starting of the simulation up to $T_{max}$, 
is characterized by the accumulation of new words 
in the agents'  memory and the establishment of correlations between them.
In this time span, the introduction of the reputation structure does not have 
any relevant role in the scaling behavior, 
which shows the same exponent ($\alpha=1.5$) as in the classical model.
The second process, which extends from $T_{max}$ to $T_{con}$, 
is characterized by the fast alignment among all memories.
This second time period ($T_{con}-T_{max}$), in contrast, is strongly influenced by the presence 
of the hierarchical structure 
introduced by  the players' reputation. In fact, if we look at its scaling 
behavior (see Figure \ref{fig_tconv}), we find a power law with an 
exponent $\beta$ slightly smaller than $1.2$.
These facts suggest that $T_{con}(P)$ can be alternative fitted by
a linear combination of these two power laws: $T_{con} = aP^{\alpha}+bP^{\beta}$.

From these consideration we can interpret 
the origin of the crossover pointed out by our previous fitting procedure.
A characteristic $P_{cha}$ value exists that defines two different scaling behaviors.
For $P<P_{cha}$ the slower power law $P^{\beta}$ is the relevant one; for $P>P_{cha}$
the relevant is  the faster one ($P^{\alpha}$).
These facts introduce a notion of small system, that can be interpreted as the 
characteristic scale for which the social structure is relevant in defining the 
overall convergence time of the  ordering transition. This happens when the 
second process of  fast alignment results in being more relevant than the 
dynamics of memory accumulation.

Finally, this analysis suggests to us an interesting supposition. 
If we arrange the agents plays in a small-world topology 
where, for the original definition of the Naming Game, 
the memory costs and, in particular, the time to reach its maximum are 
reduced ($P$-dependence \cite{asta}), 
the scaling law $T_{con}\propto P^{1.2}$ may be 
preserved for any population size.

The scaling behavior of  the convergence time that we 
obtained for relative small communities 
is quite promising. 
As far as our knowledge goes, for mean-field like interactions, 
only a complex playing strategy which introduces an ordering 
in each agent inventory (play smart strategy \cite{baronca2})
is able to obtain faster convergence times (a scaling behavior slower than $P^{1.5}$).
Such an algorithm implies 
a cognitive effort for each individual.
In contrast, our model introduces a light structure at 
the collective level of the community 
which is able to obtain similar behaviors 
for small populations. 


\section{Conclusions}

We presented results regarding a new implementation of a Naming Game.
Our model describes an open-ended system and 
embodies a hierarchical structure 
introduced by  players' reputation, which 
reflects their success in communication. 

We showed that convergence towards the use of a single word is possible.
The analysis of the scaling behavior, in dependence on the population
size, evidences that our model, in the limit of very large populations,
belongs to the same universality class of the model of Baronchelli et
al. \cite{baronca} for the behavior of the maximum number of total words, its time position and the convergence time.
In contrast, for small communities a slower $P$-dependence in convergence is found.
These results assess the robustness of the 
disorder/order transition
for real open-ended systems. They propose an 
innovative collective structure 
able to improve the system performance in reaching 
consensus, suggesting a way for optimizing 
artificial semiotic dynamics.
Moreover, we found how a dependence on the 
initial distribution of the agents' reputation 
exists, that can lead to the appearance of long 
standing quasi-stationary states characterized by a 
low numbers of words. 
From a more general point of view, 
we tested the possibility of
reaching consensus through a more realistic 
overlapping mechanism,
implemented by introducing the concept of reputation.\\ 

Finally, we want to recall how many interesting problems 
remain open and could be the subjects for future works.
First of all, it will be interesting exploring  the role 
of the system topology.
Different complex topologies  could be studied for agents 
embedded on more realistic networks.
As mentioned above,  the exploration of 
the effects of a small-world topology
on the two $P$-scaling regimes could be particularly relevant. 
In fact, if it could represent a trade-off
between the memory peak time 
and the convergence time,
a really faster convergence for large populations would appear.
Second, it could be interesting to elucidate if, for large $\sigma$ values,
it is possible to define an order/disorder transition, with a critical $\sigma$ 
value for which  consensus is not attainable.
More generally, this study should show a thorough investigation 
of the role of different $R$ distributions.



\section*{Acknowledgments} 

I am grateful to Veit Schw\"{a}mmle 
and Pietro Massignan  
for a critical reading of the manuscript and I thank the Brazilian 
agency CNPq for financial support.

\end{multicols}

\end{document}